\documentclass[preprint,12pt]{elsarticle}
 \usepackage{graphicx}

\usepackage{graphicx}

\def\slashchar#1{\setbox0=\hbox{$#1$}           
   \dimen0=\wd0                                 
   \setbox1=\hbox{/} \dimen1=\wd1               
   \ifdim\dimen0>\dimen1                        
      \rlap{\hbox to \dimen0{\hfil/\hfil}}      
      #1                                        
   \else                                        
      \rlap{\hbox to \dimen1{\hfil$#1$\hfil}}   
      /                                         
   \fi}                                         %
 
\begin{document}

\begin{frontmatter}
\title{Anomalies, gauge field topology, and the lattice}
\author{Michael Creutz}
\address{
Physics Department, Brookhaven National Laboratory\\
Upton, NY 11973, USA
}

\begin{abstract} {Motivated by the connection between gauge field topology
  and the axial anomaly in fermion currents, I suggest that the fourth
  power of the naive Dirac operator can provide a natural method to
  define a local lattice measure of topological charge.  For smooth
  gauge fields this reduces to the usual topological density.  For
  typical gauge field configurations in a numerical simulation,
  however, quantum fluctuations dominate, and the sum of this density
  over the system does not generally give an integer winding.  On
  cooling with respect to the Wilson gauge action, instanton like
  structures do emerge.  As cooling proceeds, these objects tend
  shrink and finally ``fall through the lattice.''  Modifying the
  action can block the shrinking at the expense of a loss of
  reflection positivity. The cooling procedure is highly sensitive to
  the details of the initial steps, suggesting that quantum
  fluctuations induce a small but fundamental ambiguity in the
  definition of topological susceptibility.}
\end{abstract}


\end{frontmatter}

\section{Introduction}

One of the great theoretical advances of the 1970's was the
understanding of the connection between gauge field topology and the
axial anomaly \cite{'tHooft:fv}.  One consequence of this resolution
was the realization that a CP violating parameter can be introduced
into the strong interactions.  In the context of unification, this
raised the puzzle as to why such a parameter seems to be absent or at
least experimentally quite small.  For a recent review, see
Ref.~\cite{Vicari:2008jw}.  Fujikawa \cite{Fujikawa:1979ay} provided a
simple interpretation of these phenomena in terms of the properties of
the fermionic measure in the path integral.  Indeed, this approach
provides an elegant derivation of the index theorem relating the gauge
field topology to the number of zero modes of the Dirac operator.
Witten \cite{Witten:1980sp} used large gauge group ideas and effective
Lagrangian techniques to explore the behavior of the theory on this CP
violating parameter.

There is a long history of attempts to understand this physics in the
context of lattice gauge theory, now the primary tool for
understanding non-perturbative field theory.  The idea of continuity
in space time, crucial to topology, is lost on the lattice.  With
gauge field space being a direct product of group elements on discrete
space time links, it is always possible to continuously and locally
deform fields between arbitrary configurations.  This issue can be
mollified by placing rather strong smoothness restrictions on the
differences of fields at nearby locations, thereby giving rise to a
unique continuum interpolation \cite{Luscher:1981zq}.  However these
restrictions can be shown to mutilate fundamental properties such as
reflection positivity \cite{Creutz:2004ir}.  Note that many current
simulations use discretizations that violate this condition; the point
is not that these are unacceptable in practice, but rather that any
acceptable rigorous definition of winding number must not exclude
formulations that do satisfy this axiom.

Although the issues are quite old \cite{Teper:1985rb}, they continue
to generate many recent discussions, {\it i.e.}
Refs.~\cite{Bruckmann:2006wf,Ilgenfritz:2008ia,Luscher:2010ik,Luscher:2010iy}.
Thus I return to this topic with a new definition of topology based on
a generalization of the Fujikawa picture, wherein the trace of
$\gamma_5$ times the fourth power of the Dirac operator contains a
contribution proportional to the topological charge.  The motivation
is to provide a definition that is particularly closely tied to the
fermion operator and the index theorem.

The specific lattice operator I consider is a sum over loops running
around fundamental hypercubes.  This provides a local definition of
topological charge that for smooth fields agrees with the continuum
definition.  On rough fields typical in a simulation, however, it is
not generally integer valued.  It does become so after a cooling
procedure to remove the short distance fluctuations.

Throughout I treat the fermionic fields merely as a probe of the gluon
fields.  While one could extend the discussion to include dynamical
fermions, in this paper the quarks play no role in the gluon dynamics.
Another example of using powers of the Dirac operator to probe gluonic
fields appears in \cite{Gattringer:2002gn}.  References
\cite{Horvath:2006az,Horvath:2006aj,Horvath:2006md} go further and
suggest using a Dirac operator itself to dynamically generate the
gluon interactions.

Section \ref{anomaly} reviews the argument of Fujikawa relating the
anomaly to the fermionic measure.  Section \ref{index} extends this
review to a simple derivation of the index theorem for smooth fields.
Then Section \ref{operator} introduces the specific lattice operator
under discussion and presents simulations showing that quantum
fluctuations blur any discrete topological structures.  Section
\ref{cooling} shows that such structures do become evident under the
process of cooling, also sometimes referred to as smearing, to smooth
out short distance fluctuations.  Here I also discuss the eventual
decay of the instanton structures as they shrink and fall through the
lattice due to lattice artifacts.  Section \ref{reflection} discusses
the properties imposed on this operator by reflection positivity and
the importance of the contact term present in the correlators of the
density operator.  Section \ref{sensitivity} returns to the cooling
process and discusses ambiguities in the topological charge due to a
chaotic nature of the initial cooling steps.  This raises the question
of whether topological susceptibility is actually a physical
observable, or is there a fundamental uncertainty in its definition.
I illustrate the ideas with simulations based on pure $SU(2)$ lattice
gauge theory with the standard Wilson gauge action with coupling
$\beta$ \cite{Creutz:1980zw}.  The final section summarizes the basic
ideas of the paper.

\section{Fermions and the anomaly}
\label{anomaly}

The chiral anomaly is associated with a flavor singlet transformation
on the fermion fields
\begin{eqnarray}
\psi \rightarrow e^{i\gamma_5 \phi/2}\psi\cr
\overline\psi \rightarrow \overline\psi e^{i\gamma_5 \phi/2}.\cr
\label{rotate}
\end{eqnarray}
Here $\psi$ represents the quark fields and has implicit spinor and
flavor indices.  Consider the naive kinetic term for massless quarks
\begin{equation}
\overline\psi\slashchar D\psi
=\overline\psi\gamma_\mu (\partial_\mu+ie A_\mu)\psi.
\end{equation} 
Because $\gamma_5$ anti-commutes with the anti-hermitian operator
$\slashchar D$, massless QCD naively is invariant under the above
transformation.  The essence of the chiral anomaly is that any valid
regulator must break this symmetry and leave behind physical
consequences.

A particularly elegant way to understand the connection between the
anomaly and the index theorem is due to Fujikawa
\cite{Fujikawa:1979ay}, who mapped the problem onto the fermionic
measure in the path integral.  Under the above transformation the
fermionic measure changes by
\begin{equation}
d\psi\ d\overline\psi \rightarrow \det(e^{i\gamma_5\phi})
d\psi\ d\overline\psi 
\end{equation}
Using the matrix relation $\det A = \exp({\rm Tr}\log(A))$, this
reduces to
\begin{equation}
d\psi\ d\overline\psi \rightarrow (e^{i\phi {\rm Tr}\gamma_5})
d\psi\ d\overline\psi. 
\end{equation}
If one could say that ${\rm Tr}\gamma_5=0$, the measure would be
invariant under this transformation.

The crucial result is that any valid regulator must modify the naive
tracelessness of $\gamma_5$.  The details of how this works depends on
the specifics of the regulator, but basically the issue involves
reconciling the infinity of the dimension of space time with the naive
relation ${\rm Tr}\gamma_5=0$.  A convenient approach is to use the
fermion operator itself to regulate the theory and define
\begin{equation}
{\rm Tr}\gamma_5 \equiv \lim_{\Lambda\rightarrow\infty}
\gamma_5 e^{{\slashchar D}^2/\Lambda^2}.
\end{equation}
With this it is natural to expand in the eigenvectors of
$\slashchar D$
\begin{equation}
\slashchar D |\psi_i\rangle=\lambda_i|\psi_i\rangle
\end{equation}
and define
\begin{equation}
{\rm Tr}\gamma_5=\lim_{\Lambda\rightarrow\infty}
\sum_i \langle \psi_i| \gamma_5 |\psi_i \rangle 
e^{\lambda_i^2/\Lambda^2} 
\end{equation}
Since $D$ is anti-hermitian, its eigenvalues fall on the imaginary
axis.  The eigenvectors divide into two classes, those with non-zero
$\lambda_i$ and zero modes with $\lambda_i=0$.  The former always
appear in complex conjugate pairs since $[\slashchar D,\gamma_5]_+=0$
implies $\slashchar D\gamma_5|\psi_i\rangle
=-\lambda_i\gamma_5|\psi\rangle$. Since $|\psi_i\rangle$ and
$\gamma_5|\psi_i\rangle$ have different eigenvalues, whenever
$\lambda_i\ne 0$ they are orthogonal
\begin{equation}
\langle\psi_i|\gamma_5|\psi_i\rangle=0.
\end{equation}
Such states do not contribute to the above sum for the trace of
$\gamma_5$.  Indeed, this trace only receives contributions from the
set of zero modes of the Dirac operator.  Restricted to the space
spanned by this set, $\slashchar D$ and $\gamma_5$ commute and thus
can be simultaneously diagonalized.  Let $n_+(n_-)$ be the number of
zero modes with eigenvalue $+1(-1)$ for $\gamma_5$.  Now recall
the well known index theorem that the net winding number $\nu$ of the
gauge field is given by the difference of positive and negative
chirality zero modes of the Dirac equation.
\begin{equation}
{\rm Tr}\gamma_5=n_+-n_-=\nu.
\end{equation}
Rather than being traceless, the regulated $\gamma_5$ in the path
integral is actually an operator depending on the gauge field.  On
making the chiral transformation in Eq.~(\ref{rotate}), the path
integral picks up a factor $e^{i\nu\phi}$.

Now consider the theory with a non-vanishing quark mass term
$m\overline\psi \psi$.  Under the rotation of Eq.~(\ref{rotate}), this
is not invariant but becomes
\begin{equation}
m\overline\psi \psi\rightarrow m(\cos(\phi)\overline\psi \psi
+i\sin(\phi)\overline\psi \gamma_5\psi)
\label{massrot}
\end{equation}
Since the transformation is naively a change of variables, one might
expect a theory with a mass term of the form of the right hand side of
Eq.~(\ref{massrot}) would give the same physics.  However this is
incorrect since by the above discussion this change reweights
configurations of non-trivial winding.  This gives a physically
inequivalent theory.  In particular, since the term involving
$i\overline\psi \gamma_5\psi$ is not invariant under CP, this new
theory does not respect this symmetry.

Here I have given the same rotation to each species.  With $N_f$
flavors of fermion, each contributes equally to the measure, thus the
factor appearing in the path integral is actually $e^{iN_f\nu\phi}$
and physics is periodic in $\phi$ with period $2\pi/N_f$.  This leads
to the more conventional definition of the angle $\Theta=N_f\phi$, in
which physics is periodic with period $2\pi$.  One might ask where did
the opposite chirality states go.  The are in some sense ``beyond the
cutoff,'' having been suppressed by the factor $e^{\slashchar
  D^2/\Lambda^2}$.

\section{The index theorem}
\label{index}

This approach leads to a simple derivation of the index theorem.
Assume the gauge fields are smooth and differentiable.  Writing out
the square of the Dirac operator gives
\begin{equation}
\label{dsquare}
{\slashchar D}^2=\partial^2-g^2A^2+2igA_\mu\partial_\mu+ig(\partial_\mu A_\mu)
-{g\over 2}\sigma_{\mu\nu}F_{\mu\nu}
\end{equation}
where $[\gamma_\mu,\gamma_\nu]=2i\sigma_{\mu\nu}$.  Expanding ${\rm
  Tr}\gamma_5 e^{{\slashchar D}^2/\Lambda^2}$ in powers of the gauge
field, the first non-vanishing term is contained in the fourth power
of the Dirac operator.  This involves two powers of the sigma matrices
\begin{equation}
\nu={\rm Tr}\gamma_5 e^{{\slashchar D}^2/\Lambda^2}
={1\over 2\Lambda^4}{\rm Tr}_{x,c}
e^{\partial^2/\Lambda^2}\epsilon_{\mu\nu\rho\sigma}F_{\mu\nu}F_{\rho\sigma}
\end{equation}
where ${\rm Tr}_{x,c}$ refers to the trace over space and color, the
trace over the spinor index having been done to give a factor of 4.
It is the trace over the space index that will give a divergent factor
removing the $\Lambda^{-4}$ prefactor.  Higher order terms go to zero
rapidly enough with $\Lambda$ to be ignored.

The factor $e^{\partial^2/\Lambda^2}$ serves to mollify traces over
position space.  Consider some function $f(x)$ as representing a a
diagonal matrix in position space
$M(x,x^\prime)=f(x)\delta(x-x^\prime)$ The formal trace would be ${\rm
  Tr}M=\int dx M(x,x)$, but this diverges since it involves a delta
function of zero.  Writing the delta function in terms of its Fourier
transform
\begin{equation}
e^{\partial^2/\Lambda^2} \delta(x-x^\prime)
=\int {d^4p\over (2\pi)^4} e^{ip\cdot (x-x^\prime)}
e^{-p^2/\Lambda^2}
={\Lambda^4\over 16\pi^2}e^{-(x-x^\prime)^2\Lambda^2/4}
\end{equation}
shows how this ``heat kernel'' spreads the delta function.  This
regulates the desired trace
\begin{equation}
{\rm Tr}_x f(x)\equiv
{\Lambda^4\over 16\pi^2}\int d^4x f(x).
\end{equation}
Using this to remove the spatial trace in the above gives the well
known relation
\begin{equation}
\nu={1\over 32\pi^2}{\rm Tr}_c \int d^4x
\epsilon_{\mu\nu\rho\sigma}F_{\mu\nu}F_{\rho\sigma}
={1\over 16\pi^2}{\rm Tr}_c \int d^4x
F_{\mu\nu}\tilde F_{\mu\nu}
\end{equation}
where $\tilde F_{\mu\nu}={1\over 2}
\epsilon_{\mu\nu\rho\sigma}F_{\rho\sigma}$.

\section{A local lattice operator for topology}
\label{operator}

The previous argument showed that the fourth power of the Dirac
operator has a close connection to topological charge.  To obtain
something with non-vanishing trace, one must multiply $\gamma_5$ by
each of the four space-time gamma matrices.  In the above discussion,
these factors were provided by the square of the
$\sigma_{\mu\nu}F_{\mu\nu}$ term in Eq.~(\ref{dsquare}).  

In a lattice formulation, one obtains gamma matrix factors for each
fermion hopping term, and thus can obtain the necessary factors by
hopping once in each of the four directions.  This suggests one should
study contributions to lattice operators of the form ${\rm Tr}\gamma_5
{\slashchar D}^4$.  This leaves open what lattice Dirac operator to
use.  With an overlap formulation \cite{Neuberger:1997fp}, calculating
${\rm Tr}\ \gamma_5 e^{\slashchar D^2/\Lambda^2}$ amounts to a simple
counting of zero modes.  However the overlap operator is rather
tedious to compute.  With the naive lattice discretization, the
fermion operator just involves nearest neighbor hops accompanied with
the corresponding gamma matrices.  For the free theory in momentum
space this is
\begin{equation}
\slashchar D(p) =\sum_\mu i\gamma_\mu \sin(p_\mu). 
\end{equation}
In position space this gives a factor of $\pm i\gamma_\mu/2$ for hops
in the $\pm \mu$ direction.  With gauge fields present, the hop is
accompanied by the corresponding gauge link field $U$.

Of course the naive fermion action involves doublers, but for the
above counting one could imagine removing this with a simple numerical
factor.  A complication here is that the various doublers have
different chiralities.  When some component of momentum is near $\pi$,
the slope of $\sin(p_\mu)$ in that direction brings in a minus sign,
and the doubler uses an opposite sign for that particular gamma
matrix.  Half the doublers use the opposite sign for $\gamma_5$ and
thus the naive $\gamma_5$ is actually a non-singlet operator.  As
such, the corresponding chiral symmetry is not anomalous and ${\rm
  Tr}\ \gamma_5 e^{\slashchar D^2/\Lambda^2}$ explicitly vanishes.

The simplicity of the naive Dirac operator makes it worth trying to
get around this and consider a new chiral matrix that is closer to a
flavor singlet.  In momentum space for free fermions I define
\begin{equation}
\Gamma_5=\gamma_5\prod_\mu\cos(p_\mu).
\end{equation}
Whenever a component of the momentum is near $\pi$, this introduces a
minus sign to compensate for the effective Dirac matrix used by the
doubler.  

\begin{figure}\centering
{
\includegraphics[width=4in]{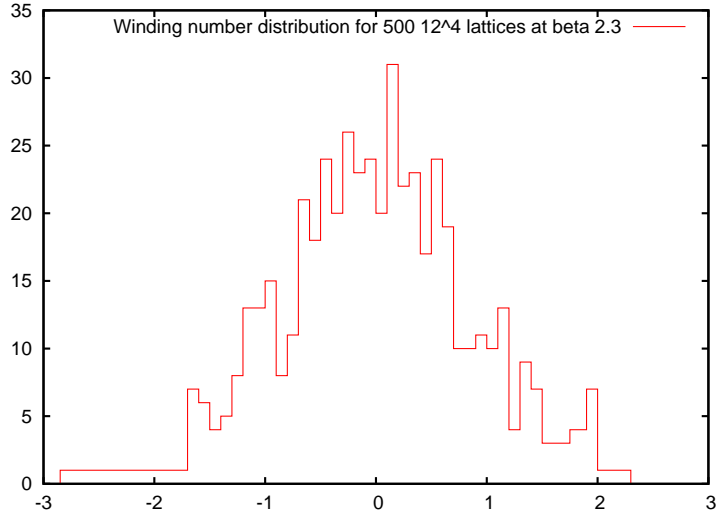}
\caption{ 
The distribution of the lattice winding number defined in the text
on a set of 500 $SU(2)$ lattices of size $12^4$ at $\beta=2.3$.
}
\label{qdistribution} 
}
\end{figure}

Going back to position space, the factors of $\cos(p_\mu)$ are
represented by averaging hoppings in the $\pm\mu$ direction.
Reintroduce the gauge fields with insertions of the link variables
into these hoppings.  Since the order of the hoppings should not
matter, consider the average over all such.  Thus our candidate for
the anomalous flavor singlet chiral matrix involves taking $\gamma_5$
augmented by a hop in each of the four space time directions.  These
hops are summed over positive and negative directions and all
orderings.  The result is that $\Gamma_5$ is a non-local operator
connecting opposite corners of lattice hypercubes.

This suggests the candidate operator for measuring topology
\begin{equation}
q\propto {\rm Tr}\ \Gamma_5 {\slashchar D}^4
\label{winding}
\end{equation}
with ${\slashchar D}$ being the naive lattice fermion operator.  Each
term in this construction involves eight hoppings, four from the
$\Gamma_5$ factor and one from each factor of ${\slashchar D}$.
Because the trace forces the hops to return to the starting site, this
quantity is gauge invariant.

\begin{figure}\centering{
\includegraphics[width=3in, angle=-90]{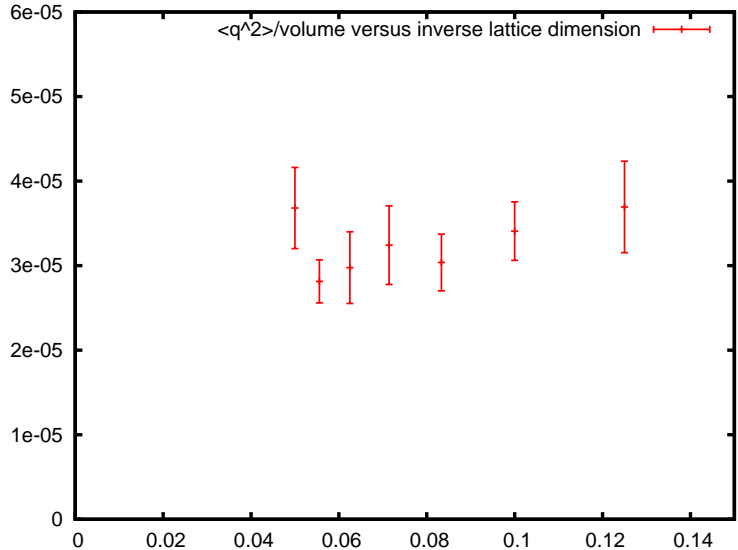}
\caption{ The average topological charge squared divided by the system
  volume versus the linear lattice size at $\beta=2.3$ for lattices of
  size $8^4$ to $20^4$.
}
\label{sizeplot} 
}
\end{figure}

So the final operator I wish to construct is defined on hypercubes.
For any given hypercube, consider all 16 directed hyper-diagonals.
For each diagonal, sum over all four-hop paths along hypercube edges
from one end to the other followed by all four-hop paths back to the
starting site.  Finally given any particular path, assign a sign
corresponding to the parity of the permutation of the initial four
hops.  This is a sum over $24^2*16$ Wilson loops.

In practice one need not actually calculate all these loops
individually.  Accumulating the forward paths into two matrices, one
being the sum and one being the sum with the sign factor included, the
desired sum of loops is immediately found from the product of the
first sum times the adjoint of the second.  Also, each diagonal need
be calculated only in one direction since the reverse is equivalent.

To normalize this construction, consider small smooth fields ask that
the sum give the classical winding number
\begin{equation}
q\rightarrow \nu={1\over 16\pi^2}\int d^4x {\rm Tr}_c F\tilde F
\end{equation}
where ${\rm Tr}_c$ refers to a trace over color matrices.  The result
is that one should multiply the above sum over Wilson loops by a
factor of ${1\over 3*2^{10} \pi^2}$.  Note that this is independent of
the size of the gauge group.

Because the lattice fields tend to be quite rough, this quantity has
no need to peak at integer values.  In Fig. (\ref{qdistribution}) I
show the distribution of $q$ over a set of 500 independent gauge
configurations with gauge group $SU(2)$ at a coupling $\beta=2.3$ on a
$12^4$ site lattice.

The average value of $q^2$ should scale with the lattice volume.  In
Fig.~(\ref{sizeplot}) I show this quantity calculated at $\beta=2.3$
on lattices of size from $8^4$ to $20^4$.  On the $18^4$ lattice, the
topological susceptibility per unit volume in lattice units is
$(2.8\pm 0.25)\times 10^{-5}$.

\begin{figure}\centering{
\includegraphics[width=3in, angle=-90]{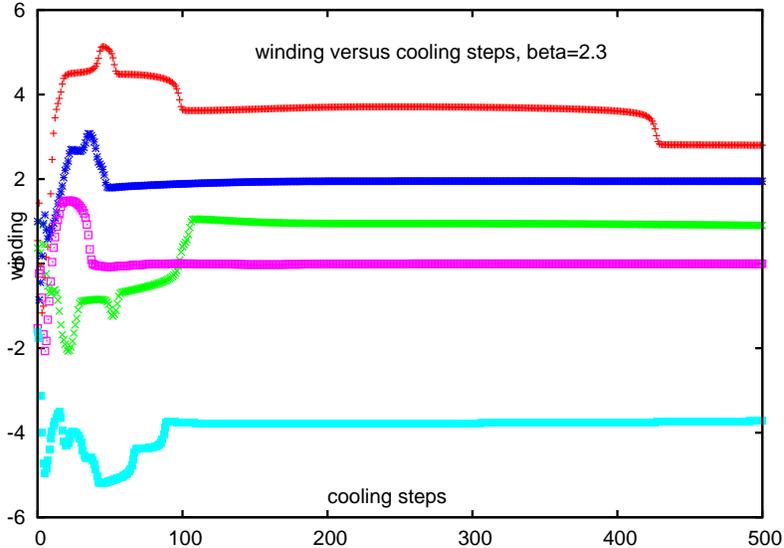}
\caption{ The winding number as a function of cooling steps for a set
  of 5 lattices of size $16^4$ at $\beta=2.3$. Note how it settles
  into approximately integer values with occasional jumps between
  different windings.  }
\label{windingcooling} 
}
\end{figure}

\section{Cooling}
\label{cooling}

It has been known for some time that to expose topological structures
in lattice configurations requires some cooling procedure to remove
short distance roughness \cite{Teper:1985rb}.  A particularly simple
process is to loop over the lattice in a checkerboard manner and
replace each link with the group element which minimizes the total
action of the six plaquettes attached to that link.  This can be
modified by over or under relaxation, which I will discuss later.
Such processes monotonically decrease the total action. In
Fig.~(\ref{windingcooling}) I show how the above winding number
evolves for 5 independent $16^4$ site lattices at $\beta=2.3$.  Note
how the winding tends to fall into discrete values, with jumps between
them as the cooling proceeds further.  Indeed, with long enough
cooling the winding always appears to eventually drop to zero.  Also
note that the discrete values tend to be somewhat below integers; this
is presumably a lattice artifact which becomes more severe at higher
winding number.  Finally, note that the initial relaxation steps
appear to be rather chaotic.

\begin{figure}\centering{
\includegraphics[width=3in, angle=-90]{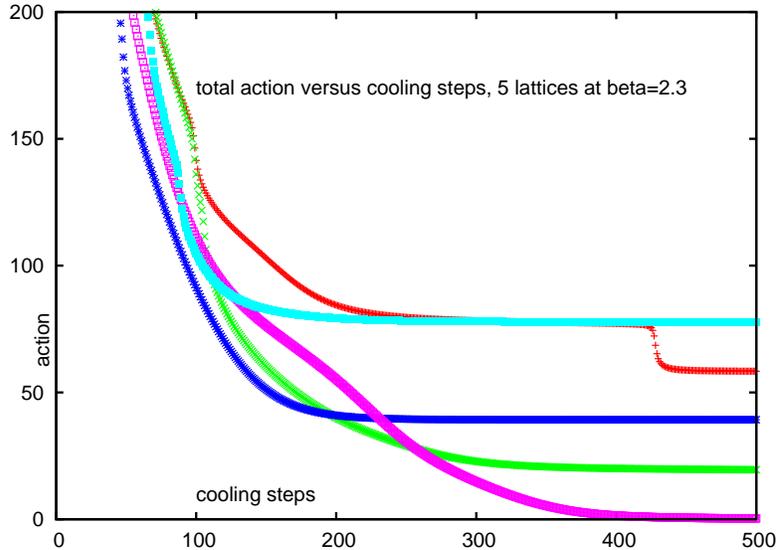}
\caption{ The total Wilson action as a function of cooling steps for a
  set of 5 lattices of size $16^4$ at $\beta=2.3$.  The level portions
  occur at approximate multiples of the classical instanton action of
  $2\pi^2$.  These are the same lattices as in
  Fig.~\ref{windingcooling}.}  
\label{actioncooling} 
}
\end{figure}

The discrete levels of non-trivial winding can also be seen directly
in the total action.  In Fig.~(\ref{actioncooling}) I show the
evolution of the action for cooling the same 5 configurations.  The
classical instanton result for the lattice action is that these levels
should appear at a multiple of $2\pi^2$, which seems to be well
satisfied.

\begin{figure}\centering{
\includegraphics[width=3in, angle=-90]{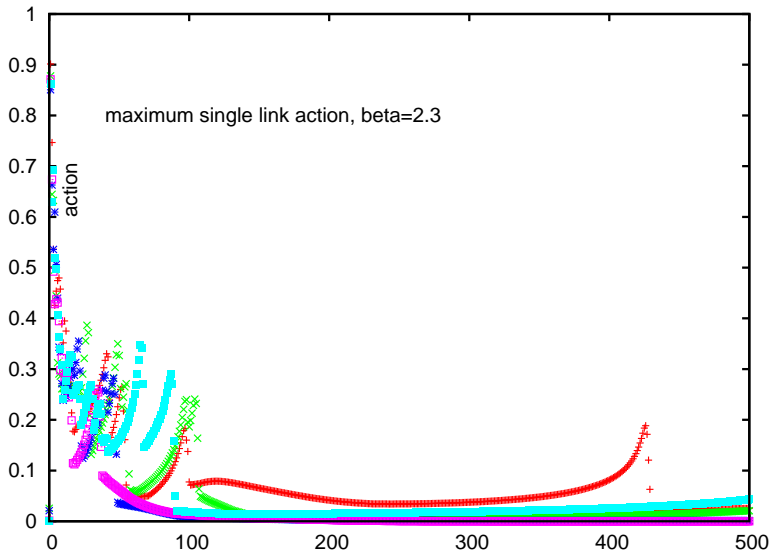}
\caption{ The maximum Wilson action associated with a single lattice
  link as a function of cooling steps for a set of 5 lattices of size
  $16^4$ at $\beta=2.3$.  The peaks represent ``instantons'' falling
  through the lattice.}
\label{coolingpeaks} 
}
\end{figure}

To better understand the jumps between levels, it is interesting to
look at the maximum local action associated with a single link.  This
is plotted in Fig.~(\ref{coolingpeaks}) for the same 5 configurations.
Note the evident peaks at the points where the winding jumps.
Although the total action decreases monotonically, this is not in
general true locally.  This is because the instanton like structures
can decrease in size with their action being concentrated in a smaller
region.  Eventually the structures drop down to the lattice size and
collapse ``through the lattice.''  This shrinking of topological
structures is a lattice artifact.  In particular, the classical
continuum theory is conformally invariant with an instanton action
independent of size.

The height of the observed peaks is approximately 0.2.  As each link
involves six plaquettes, one can stop the decay by forbidding each
plaquette to be larger than one sixth of that, or something like 0.03.
This compares well with the ``admissibility condition'' of Luscher
\cite{Luscher:1981zq} which, when obeyed, allows the gauge fields to
be uniquely continued between lattice sites to form a smooth continuum
field.  Unfortunately, such a condition has been shown to be
inconsistent with reflection positivity \cite{Creutz:2004ir}.

\section{The decay of an instanton}

It is perhaps interesting to study the way winding number disappears
with cooling in a more controlled manner.  To do this I construct a
configuration that quickly settles into a classical instanton and then
watch its evolution under cooling.  For an initial state I start with
all links unity and then do a gauge transformation with gauge function
\begin{equation}
h(x_\mu)={x_0+i\vec x \cdot \vec \sigma\over \sqrt{x_0^2+\vec x^2}}
\end{equation}
Here I define $x_\mu$ as the distance of a given site from the center
of a hypercube at the center of the lattice.  Since this is just a
gauge transformation, this leaves a configuration which still has
vanishing action.  As one moves away from the lattice center the links
approach unity everywhere except for those links crossing the
boundary.  I now make the action non-trivial by replacing all links
that cross the boundary with unity.  With this starting configuration,
I then apply the cooling process.

\begin{figure}\centering{
\includegraphics[width=3in, angle=-90]{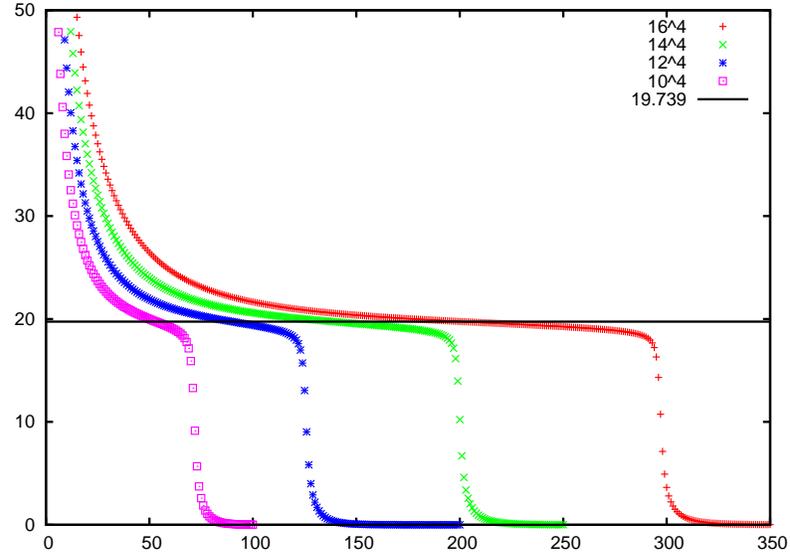}
\caption{The evolution of the action under cooling with the initial
  configuration described in the text on lattices sizes from $10^4$ to
  $16^4$.  The classical instanton action is marked by the horizontal
  line.  }
\label{decayaction} 
}
\end{figure}

\begin{figure}\centering{

\includegraphics[width=3in, angle=-90]{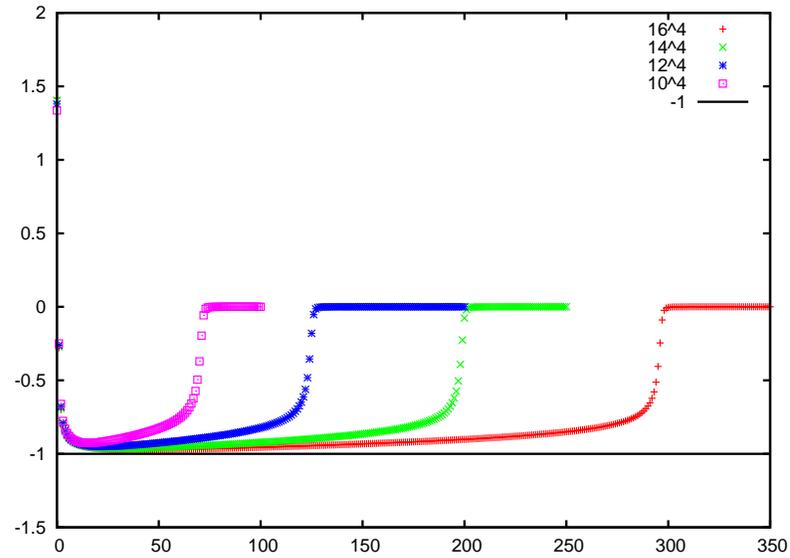}
\caption{The evolution of the winding number under cooling with the initial
  configurations described in the text.  The classical anti-instanton winding
  is marked by the horizontal line.
}
\label{decaywinding} 
}
\end{figure}

\begin{figure}\centering{

\includegraphics[width=3in, angle=-90]{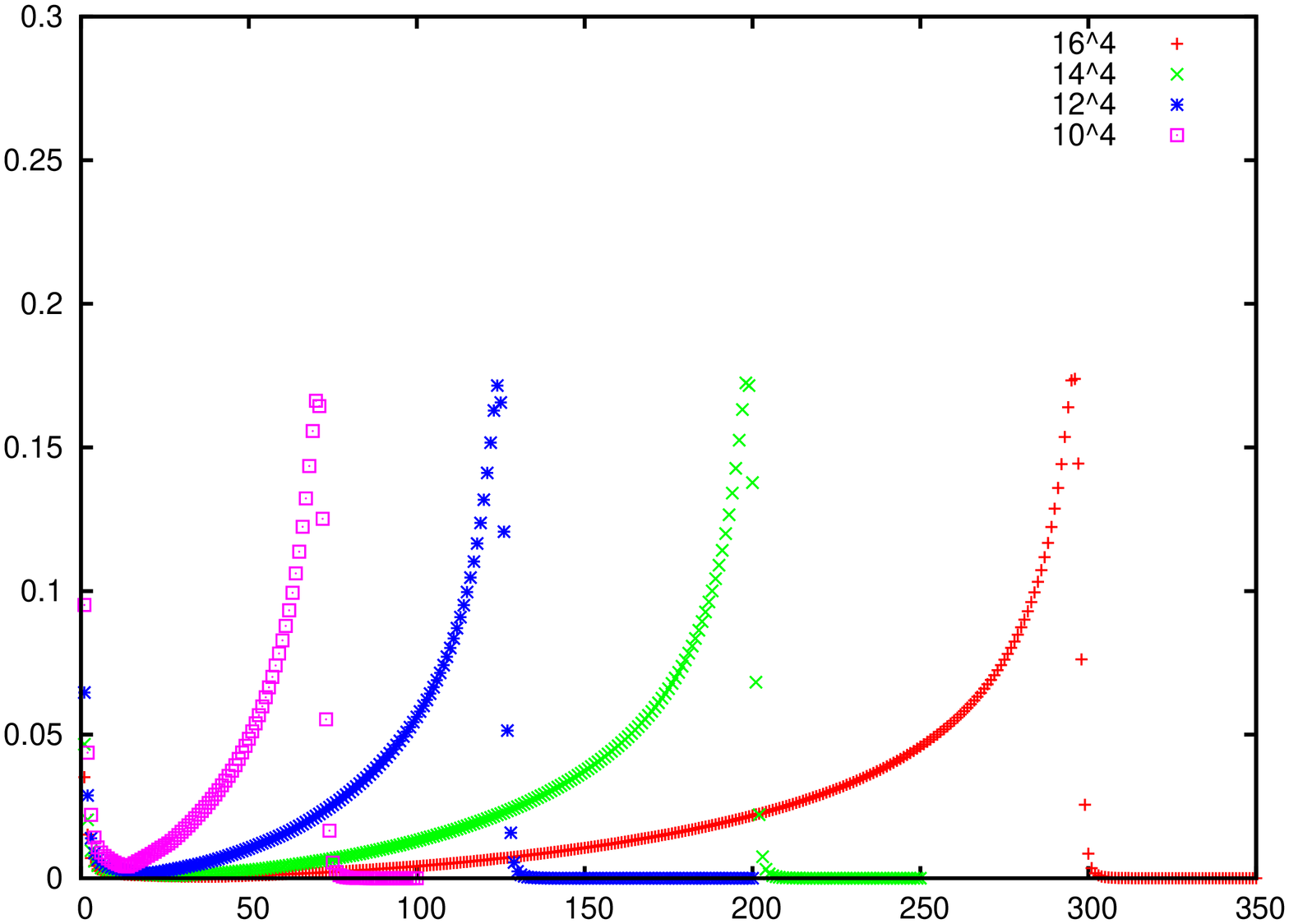}
\caption{The evolution of the largest action associated with a single
  link under cooling with the initial
  configurations described in the text.  The peak represents the
  instanton collapsing through the lattice.
}
\label{decaypeaks} 
}
\end{figure}

In Fig.~(\ref{decayaction}) I show the evolution of the action for
such configurations under cooling for various size lattices.  In the
figure I also indicate $2\pi^2$, which is the action of a smooth
classical instanton.  Note how the action quickly relaxes to this
value, then gradually decreases while the instanton shrinks and
finally drops to zero.  The number of steps for this collapse
increases with lattice size because the initial topology becomes
spread over a larger region.

Fig.~(\ref{decaywinding}) shows the topological winding number during
this process.  After the roughness at the boundary smooths out, this
quickly drops to near $-1$, indicating that the construction actually
gives rise to an anti-instanton.  The evolution then remains near
unity until the time of the action drop.

Finally, in Fig.~(\ref{decaypeaks}) I plot the maximum action
associated with a single link during this cooling.  As the instanton
becomes smaller, the action is more concentrated and a peak appears
just before the collapse.  This is an analogous peak to those
appearing in the cooling of equilibrated lattices as in
Fig.~(\ref{coolingpeaks}).

\section{Reflection positivity and topological charge}
\label{reflection}

\begin{figure}\centering{
\includegraphics[width=3in, angle=-90]{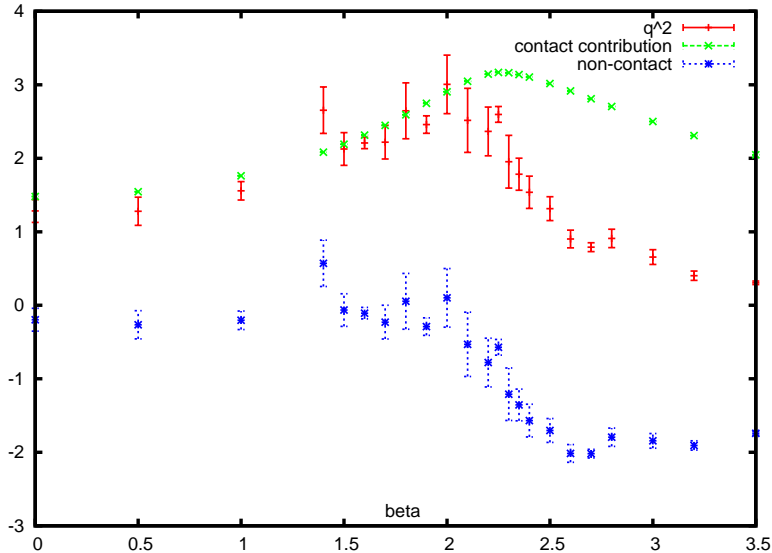}
\caption{ The average topological charge squared for $SU(2)$ on a
  $16^4$ lattice versus $\beta$ plotted along with the contact and
  non-contact contributions separated.  As required by reflection
  positivity, the non-contact contribution is negative.  }
\label{q2fig} 
}
\end{figure}

Reference \cite{Seiler:2001je} pointed out an interesting property
that any local definition of topological charge must have.  Because
the underlying operator $F\tilde F$ is odd under time reversal,
reflection positivity requires its correlator with itself at
non-vanishing separation to be negative.  On the other hand, the
square of its integral over space time must be positive.  Therefore in
calculating the square of the topological charge, there is a subtle
cancellation between the positive contact term and the negative
contribution from correlations at non-vanishing separations.  This is
a somewhat non-intuitive result that led Horvath
\cite{Horvath:2003yj,Horvath:2005cv} to interpret the typical vacuum
state in terms of a highly crumpled field structure.  The separation
of the topological charge into contact and non-contact pieces is shown
in Fig.~(\ref{q2fig}).

\begin{figure}\centering{
\includegraphics[width=3in, angle=-90]{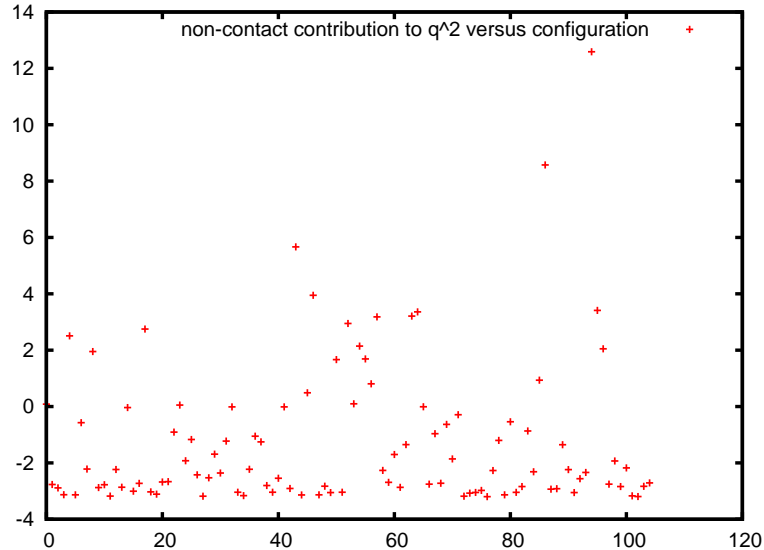}
\caption{ The non-contact contribution to $q^2$ over a sequence of 105
  gauge configurations.  Although the average is negative as required,
  there are large fluctuations giving a positive contribution from
  specific configurations.  This figure is using $SU(2)$ on a $16^4$
  lattice at $\beta=2.3$.}
\label{noncontact} 
}
\end{figure}

The negative nature of the correlators of topological charge is a
statement about their average.  It is, however, possible for
fluctuations to give these correlators positive contributions on a
configuration by configuration basis.  In Fig.~(\ref{noncontact}) I
plot the evolution of the non-contact term over a sequence of 105
configurations on a $16^4$ lattice at $\beta=2.3$.

\begin{figure}\centering{
\includegraphics[width=3in, angle=-90]{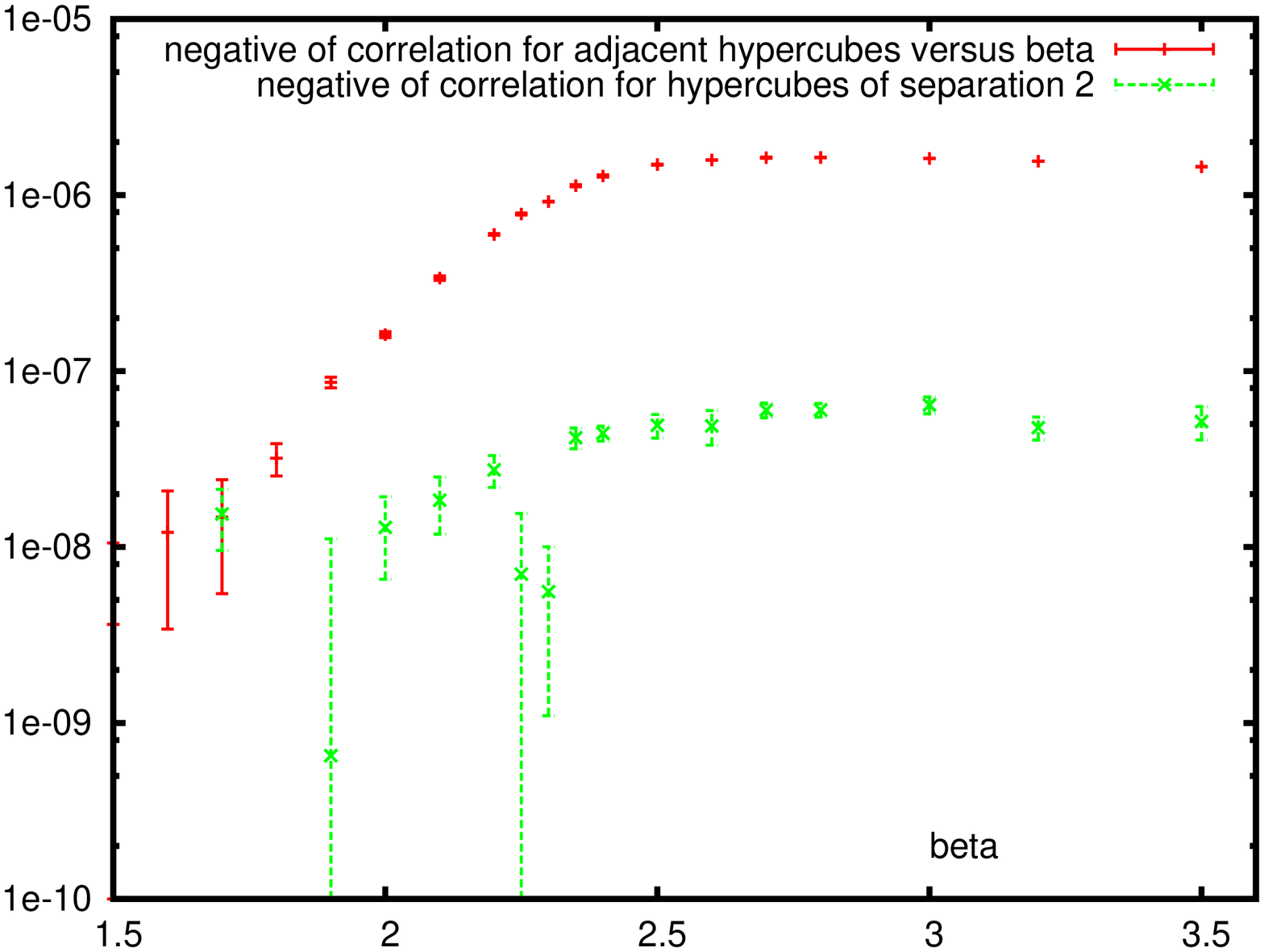}
\caption{ The negative of the correlation of winding number density
  between adjacent hypercubes and hypercubes of on axis separation two
  plotted as a function of the coupling beta.  As required by
  reflection positivity, the correlation is negative.  This plot is
  for $SU(2)$ gauge theory on a $16^4$ lattice.  }
\label{plotcor} 
}
\end{figure}

The definition of charge density used here is highly local, involving
single hypercubes.  As a consequence, the negative nature of the
correlation starts immediately for adjacent hypercubes.  In
Fig.~(\ref{plotcor}) I show the correlation between the charge density
on hypercubes separated along an axis by one and two lattice units.

The cooling process discussed in Section \ref{cooling} smooths out the
fields in a way that does not maintain the negative nature of this
correlation.  Indeed, during the cooling process the non-contact part
quickly comes to dominate the total contribution to the square of the
topological charge. 

At long distances the correlator between charge densities should fall
exponentially with the mass of the lightest singlet pseudoscalar
meson.  For pure gauge fields, this is a glueball, but with quarks of
physical masses present this will be the eta prime meson.  (More
precisely, with physical quark masses the eta prime decays into three
pions which give the ultimate asymptotic behavior; I ignore this
complication here.)  Thus measuring this correlator provides a route
to the eta prime mass which replaces the complexity of separating
connected and disconnected quark diagrams with potentially large
statistical fluctuations.

\section{Sensitivity to initial conditions}
\label{sensitivity}

The charge operator defined above does not give an integer value on a
typical gauge configuration.  Indeed, some cooling is necessary to
remove short distance fluctuations before discrete winding numbers are
observed.  Empirically with enough cooling any $SU(2)$ gauge
configuration appears to eventually decay to a state of zero action,
gauge equivalent to the vacuum.  This is not in general true for
larger gauge groups.  Indeed, for $SU(N)$ with $N\ge 5$ there exist
non-trivial gauge configurations that are stable minima under
arbitrary small variations of the fields but have nothing to do with
topology.  For example, consider all links to be unity except for a
few isolated links with values in the group center nearest the
identity, {\rm i.e.} of form $e^{\pm 2\pi i/N}$.  When $N$ is five or
more, small deviations from these center elements will raise the
action.  Such configurations do not involve gauge field winding, but
are sufficient to give stable non-minimal action states.

Since topological configurations appear to always eventually cool to
triviality, using cooling to define winding number requires an
arbitrary selection for cooling time.  Modifying the Wilson action can
prevent the winding decay.  As discussed earlier, on forbidding the
lattice action on any given plaquette from becoming larger than a
small enough number, the peaks seen in Fig~(\ref{coolingpeaks}) cannot
be crossed.  Such a condition, however, violates reflection positivity
and arbitrarily selects a special instanton size where the action is
minimum.

\begin{figure}\centering{
\includegraphics[width=3in, angle=-90]{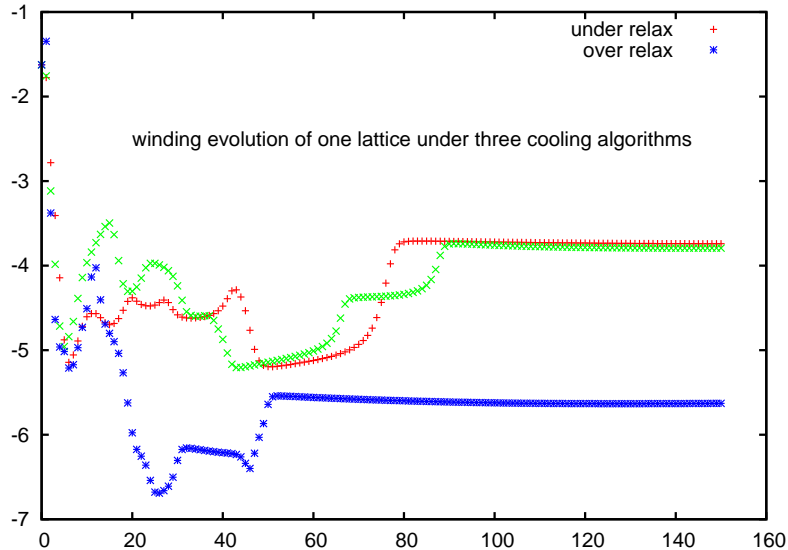}
\caption{ The topological charge evolution for three different cooling
  algorithms on a single $\beta=2.3$ lattice configuration for $SU(2)$
  on a $16^4$ lattice.  }
\label{relaxparm} 
}
\end{figure}

Cooling time is not the only issue here.  While integer winding does
not appear before cooling, note from Fig.~(\ref{windingcooling}) that
the initial cooling stages seem quite chaotic.  This raises the
question of whether the discrete stages reached after some cooling
might depend rather sensitively on the cooling algorithm.  In
Fig.~(\ref{relaxparm}) I show the evolution of a single lattice with
three different relaxation algorithms.  This lattice is the one in
Fig.~(\ref{windingcooling}) with the most negative winding after 50
cooling steps.  One algorithm is the same as used in
Fig.~(\ref{windingcooling}) where the links are replaced using
checkerboard ordering with the element that minimizes the action
associated with that link.  This is done by projecting the sum of
staples that interact with the link onto the group.  For the second
approach, an under-relaxed algorithm adds the old link to the sum of
the neighborhood staples before projecting onto the new group element.
Finally, an over-relaxed approach subtracts the old element from the
staple sum.  The resulting windings not only depend on cooling time,
but also on the specific algorithm chosen.
 
In an extensive analysis, Ref.~\cite {Bruckmann:2006wf} has compared a
variety of filtering methods to expose topological structures in gauge
configurations.  All schemes have some ambiguities, but when the
topological structures are clear, the various approaches when
carefully tuned give similar results.  Nevertheless the question
remains of whether there is a rigorous and unambiguous definition of
topology that applies to all typical configurations arising in a
simulation.  Luscher has recently discussed using a differential flow
with the Wilson gauge action to accomplish the cooling
\cite{Luscher:2010iy}.  This corresponds to the limit of maximal
under-relaxation.  This approach presumably still allows the above
topology collapse unless prevented by something like the admissibility
condition or the selection of an arbitrary flow time.  In addition, if
one wishes to determine the topological charge of a configuration
obtained in some large scale dynamical simulation, it is unclear why
one should take the particular choice of the Wilson gauge action.

The high sensitivity to the cooling algorithm on rough gauge
configurations suggests that there may be an inherent ambiguity in
defining the topological charge of typical gauge configurations and
consequently a small ambiguity in the definition of topological
susceptibility.  It also raises the question of how smooth is a given
definition as the gauge fields vary; how much correlation is there
between nearby gauge configurations?  Although such issues are quite
old \cite{Teper:1985rb}, they continue to be of considerable interest
\cite{Bruckmann:2009cv,Moran:2010rn,Luscher:2004fu}.

As topological charge is suppressed by light dynamical quarks, this is
connected to whether the concept of a single massless quark is well
defined \cite{Creutz:2003xc}.  Dynamical quarks are expected to
suppress topological structures, and the chiral limit with multiple
massless quarks should give zero topological susceptibility with a
chiral fermion operator, such as the overlap.  However, with only a
single light quark, the lack of chiral symmetry indicates that there
is no physical singularity in the continuum theory as this mass passes
through zero.  Any scheme dependent ambiguity in defining the quark
mass would then carry through to the topological susceptibility.

One might argue that the overlap operator solves this problem by
defining the winding number as the number of zero eigenvalues of this
quantity.  Indeed, it has been shown
\cite{Giusti:2004qd,Luscher:2004fu} that this definition gives a
finite result in the continuum limit.  As one is using the fermion
operator only as a probe of the gluon fields, this definition can be
reformulated directly in terms of the underlying Wilson operator
\cite{Luscher:2010ik}.  While the result may have a finite continuum
limit, the overlap operator is not unique.  In particular it depends
on the initial Dirac operator being projected onto the overlap circle.
For the conventional Wilson kernel, there is a dependence on a
parameter commonly referred to as the domain-wall height.  Whether
there is an ambiguity in the index defined this way depends on the
density of real eigenvalues of the kernel in the vicinity of the point
from which the projection is taken.  Numerical evidence
\cite{Edwards:1998wx} suggests that this density decreases with
lattice spacing, but it is unclear if this decrease is rapid enough to
give a unique susceptibility in the continuum limit.  The
admissibility condition also successfully eliminates this ambiguity;
however, as mentioned earlier, this condition is inconsistent with
reflection positivity.

Whether topological susceptibility is well defined or not seems to
have no particular phenomenological consequences.  Indeed, this is not
a quantity directly measured in any scattering experiment.  It is only
defined in the context of a technical definition in a particular
non-perturbative simulation.  Different valid schemes for regulating
the theory might well come up with different values; it is only
physical quantities such as hadronic masses that must match between
approaches.  The famous Witten-Veneziano relation
\cite{Witten:1979vv,Veneziano:1979ec} does connect topological
susceptibility of the pure gauge theory in the large number of color
limit with the eta prime mass.  The latter, of course, remains well
defined in the physical case of three colors, but the finite $N_c$
corrections to topology can depend delicately on gauge field
fluctuations, which are the concern here.

\section{Conclusions}

I have discussed a particular lattice definition of topological charge
density on the lattice.  This is motivated by the index theorem
relating the zero modes of the Dirac operator to the winding of the
gauge field.  Quantum fluctuations connected with rough gauge fields
generally give a non-integer value to the overall charge.  Cooling
schemes can remove these fluctuations, giving an integer value to the
index.  However the specifics of the final winding depend on details
of the cooling procedure.  Furthermore, on extensive cooling the
topological structures eventually shrink and collapse through the
lattice. These behaviors raise the question of whether there is a
fundamental scheme dependent ambiguity in the definition of
topological susceptibility.

\section*{Acknowledgements}

 I am grateful to the Alexander von Humboldt Foundation for supporting
 visits to the University of Mainz where part of this study was
 carried out.  This manuscript has been authored by employees of
 Brookhaven Science Associates, LLC under Contract
 No. DE-AC02-98CH10886 with the U.S. Department of Energy. The
 publisher by accepting the manuscript for publication acknowledges
 that the United States Government retains a non-exclusive, paid-up,
 irrevocable, world-wide license to publish or reproduce the published
 form of this manuscript, or allow others to do so, for United States
 Government purposes.

\end{document}